\documentclass[a4paper,12pt]{article}
\pdfoutput=1
\usepackage[top=1in,bottom=1in,left=0.5in,right=0.5in]{geometry}

\usepackage[numbers]{natbib}
\usepackage{soul}
\usepackage{caption}
\usepackage{graphicx,caption,xcolor}
\usepackage{blindtext}
\usepackage{authblk}

\begin{document}

\def\floatpagepagefraction{1}
\def\textpagefraction{.001}
%
\title{\bf In situ ultrasound imaging of shear shock waves in the porcine brain}                      
%


\author[1,2]{Sandhya Chandrasekaran}
\author[1]{Francisco Santibanez}
\author[3]{Bharat B. Tripathi}
\author[1]{Ryan DeRuiter}
\author[1]{Gianmarco F. Pinton \thanks{Corresponding author, gia@email.unc.edu}}

\affil[1]{\normalsize Joint Department of Biomedical Engineering, University of North Carolina at Chapel Hill and North Carolina State University, 9212A Mary Ellen Jones Building, 116 Manning drive, Chapel Hill,  North Carolina, USA.}
\affil[2]{\normalsize Department of Mechanical Engineering, North Carolina State University, Raleigh, North Carolina, USA.}
\affil[3]{\normalsize School of Mathematics, Statistics and Applied Mathematics, NUI Galway, University Road, Galway, Ireland.} 
\date{}

\maketitle

\begin{abstract}
Using high frame-rate ultrasound and high sensitivity motion tracking, we recently showed that shear waves sent to the \textit{ex vivo} porcine brain develop into shear shock waves with destructive local accelerations inside the brain which may be a key mechanism behind deep traumatic brain injuries. 
Direct measurement of brain motion at an adequate frame-rate during impacts has been a persistent challenge. Here we present the ultrasound observation of shear shock waves in the acoustically challenging environment of the \textit{in situ} porcine brain during a single-shot impact.
The brain was attached to a plate source which was vibrated at a moderate amplitude of 25$g$, to propagate a 40~Hz shear wave into the brain. Simultaneously, images of the moving brain were acquired at 2193 images/s, using a custom imaging sequence with 8 interleaved ultrasound transmit-receive events, designed to accurately track shear shocks. 
To achieve a long field-of-view, wide-beam emissions were designed using time-reversal ultrasound simulations and no compounding was used to avoid motion blurring.
A peak acceleration of 102$g$ was measured at the shock-front, 7.1~mm deep inside the brain. It is also shown that experimental shear velocity, acceleration, and strain-rate waveforms in brain are in excellent agreement with theoretical predictions from a custom higher-order finite volume method hence demonstrating the capabilities to measure rapid brain motion even in the presence of strong acoustical reverberations from the porcine skull.

\end{abstract}

\maketitle
\section{Introduction}

The exact relationship between local brain motion and traumatic brain injury (TBI) remains  poorly understood in the large strain and high rate regime, due to challenges of obtaining direct and quantitative measurements of brain motion during injurious impacts. TBI studies typically rely on acceleration measurements of the head \citep{rimel1981disability,greenwald2008head,Campolettano2019,Canrillo2015,CRIPTON20141,Daniel2012,Rowson2013}. However, local brain motion under impact is far more complex than the overall rigid motion of the skull \citep{bayly2005deformation,sabet2008deformation}. 

Since injurious experiments on the living brain are infeasible due to ethical reasons and since tissue degrades rapidly post mortem, TBI studies rely extensively on simulations \citep{Miller2019,zhang2004proposed,ji2014head,kleiven2007}. Finite element-based models for instance, use material properties of brain samples where the elastic constants vary by 2-3 orders of magnitude \citep{goriely2015mechanics}. There is therefore a need to directly measure local brain motion during impact to establish its link with injury and in order to validate TBI simulations.

Shear shocks are fundamentally unlike acoustical shocks which have been studied extensively including within the context of TBI. The soft tissue in the brain has nonlinear shear properties that are several orders of magnitude larger than its compressional properties. A typical Mach number (particle-velocity/wave-speed) for compressional waves in soft tissue is on the order of $10^{-4}$, and on the order of 1 for shear waves \citep{pinton2010nonlinear,Espindola2017}. This is due to the very low value of the shear wave velocity ($\sim$2 m/s) which in the case of a violent impact is comparable to the particle velocity (typically 2 m/s
or higher).  Consequently these extremely nonlinear shear waves can generate shock fronts within a single propagation wavelength i.e. areas with smooth shear waves can be adjacent to areas with violent shear shocks. 

Impacts typically last $<$50~ms \citep{cloots2013multi} during which, the brain undergoes large, nonlinear strains (0.07) at high rates (40~$s^{-1}$) \citep{Sullivan2015}, which are challenging to sample adequately using conventional imaging techniques.
X-rays can image the human brain at high frame-rates of 10,000 images/s \citep{hardy2007study}, but are limited by poor soft tissue contrast. Although high speed MRI offers high resolution, the frame-rates are too low (167 images/s \citep{sabet2008deformation}) to sample brain motion at rapid time scales $<$50~ms. The high compressional wave speed of soft tissue compared to the shear wave speed enables the use of acoustics to measure shear wave dynamics in the brain \citep{catheline2003observation}. 
Recently, using high frame-rate ultrasound (at 6000 images/s) and high sensitivity displacement tracking ($<$ 1~$\rm \mu m$ ), we showed that, planar shear waves propagated in the \textit{ex vivo} porcine brain develop into destructive shear shock waves \citep{Espindola2017}. The 35$g$ excitation wave developed into a 300$g$ shock-front deep inside the brain which we hypothesize, may be responsible for certain types of TBI. In the past studies, the imaging was performed on \textit{ex vivo} porcine brains embedded in gelatin, which presented a) fewer sources of image degradation as compared to the reflective skull surfaces around the \textit{in situ} brain and b) an indirect mechanism of transmitting shear waves into the brain. In this previous work, a flash focus imaging sequence was designed to improve tracking performance as compared to  plane-wave compounding approaches. Multiple focused ultrasound propagation events were used to image partial sections of the \textit{ex vivo} brain during each shear impact, to maintain a high frame-rate as well as high signal-to-noise ratio (SNR) \citep{espindola2017flashfocus}. However, this approach required multiple impacts (i.e. 16) to be generated in the brain, limiting our potential to investigate injury in the live brain in the realistic, single impact regime. 

Here we develop an imaging method that a) measures shear shock wave formation in the challenging \textit{in situ} environment presented by the skull b) requires only one impact and c) uses a shear wave source that is directly attached to the brain. A high frame-rate ultrasound imaging sequence (2193 images/s) was designed to interleave multiple transmit-receive events to image different sections of the brain but collectively synchronize with a single mechanical shake i.e., only a single impact is necessary to quantitatively image brain motion rather than multiple impacts. A wide-beam emission was adopted for a long field of view and to optimize the number of ultrasound  events necessary to maintain a high SNR throughout the image width. The wide emission pulse was designed using a finite difference time domain-based ultrasound simulation tool (referred to as ``Fullwave") that we have previously developed \citep{pinton2009heterogeneous,pinton2011erratum,pinton2014spatial}, in conjunction with a time-reversal technique \citep{pinton2011effects,Fink1992,pernot2007vivo,soulioti2019super,pinton2012direct}. 

Brain displacements were measured from the beamformed  radio-frequency (RF) data using an adaptive Normalized Cross Correlation (NCC)-based tracking algorithm that has been previously validated for shear shock wave motion \citep{pinton2014adaptive}. The experimentally detected shear shock waves were characterized in terms of particle velocity, acceleration, strain-rate, and amplification ratio and were compared to theoretical predictions of nonlinear shear wave propagation in a viscoelastic brain medium, using a Piecewise-Parabolic-Method based simulation tool \citep{Tripathi2019_PPM2D}. 

\begin{figure}[t]
\centering
\includegraphics[width=0.659\textwidth,trim=0 0 0 0,clip]{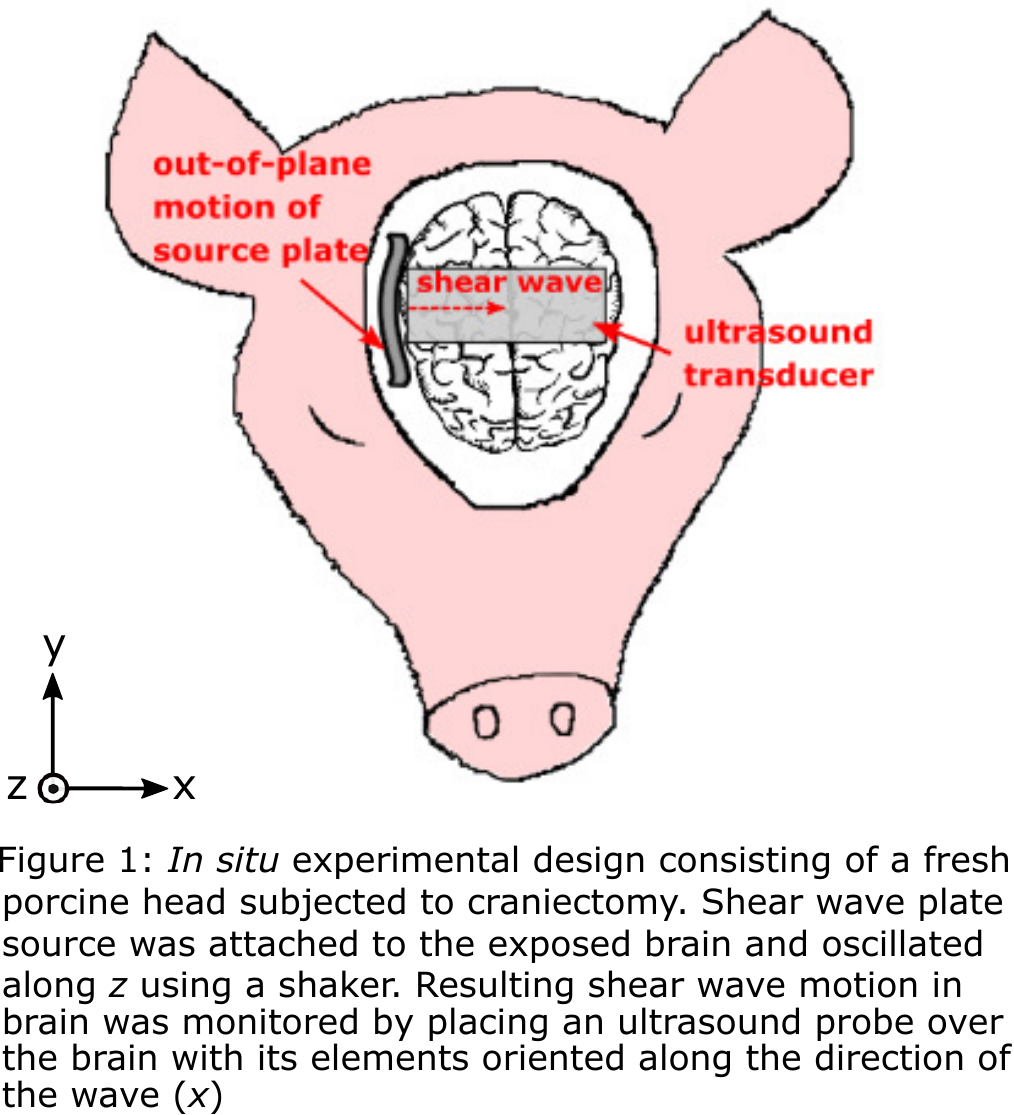}
\end{figure}
%
%
%
%

\section{Methods}

Porcine heads were extracted 20 minutes post-mortem from 4 month old Yorkshire pigs (weighing $\sim$70~lbs). A craniectomy and durectomy was performed a) to directly attach the \textit{\textit{in situ}} brain to an external excitation and b) to provide access to the brain for ultrasound imaging. The head was secured in place using a stereotactic frame. A 3D-printed plate conforming to the brain surface (circular arc cross section, 28~mm radius) was adhered with cyanoacrylate to one hemisphere of the brain (Fig.1) and vibrated along the $z$ axis by coupling to an electromechanical shaker (Vibration Test Systems, Aurora, Ohio, USA), which propagated shear waves in the brain along $x$.  The ultrasound imaging was performed using a research scanner (Verasonics, Kirkland, Washington, USA) by placing a 7.8 MHz ultrasound probe (ATL L11-5) $\sim$6~mm above the brain with its lateral axis oriented along $x$, over a layer of ultrasound gel for acoustical coupling.

\subsection{Shear wave propagation in brain using a mechanical shaker}

To impart shear impacts to the brain, a 10 cycle, 40~Hz sinusoidal train was enveloped using a -80 dB Chebychev window, amplified and sent to the shaker. The shaker response was recorded using an attached linear accelerometer (PCB Piezotronics, Inc., Depew, NY, USA) (Fig. 2). Although other excitations including non-oscillatory ones are possible, this pulse was chosen to have control over the spectral content and to simplify the analysis of the shock wave physics.  A 40~Hz impact was chosen from the range of frequencies (10-100~Hz) in head impact responses measured using an instrumented mouth-guard in various contact sports \citep{laksari2015resonance,hernandez2015six}. Moreover at this frequency, the shear wavelength is 5.4~cm, allowing the shock formation distance to be well within the brain for accessibility to imaging.

\begin{figure}[h]
\vspace{20mm}
\setlength{\unitlength}{\textwidth}
\begin{picture}(0.2,0.17)(0.27,0.051)
\put(0.50,-0.){\includegraphics[width=0.67\textwidth,trim=0 0 0 5,clip]{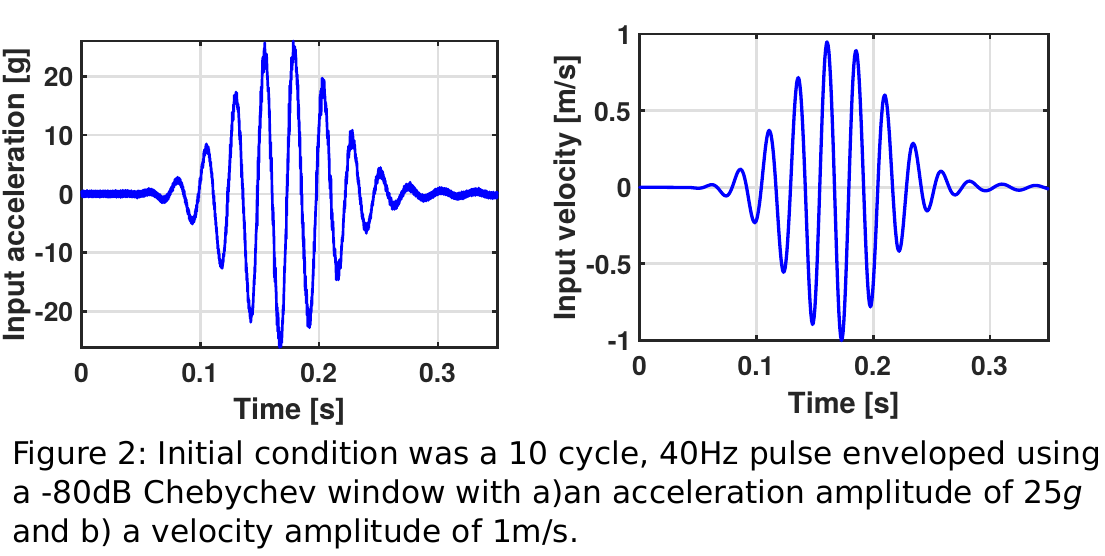}}
\put(0.77,0.15){\scriptsize \color{black}{{\bf (a)}}}
\put(1.1,0.15){\scriptsize \color{black}{\bf {(b)}}}
\end{picture}
\label{fig:input}
\end{figure}

%

\subsection{Ultrasound imaging and motion tracking}
Brain motion was captured using a custom imaging sequence with 8 transmit-receive events focused at different lateral locations (Fig.3(a)).  The 8 events were interleaved to obtain one image of the brain in motion, i.e. firing ordering was at locations 1,5,2,6,3,7,4,8. Subsequently, the next set of 8 events generated the next image of the moving brain and in this manner, 4560 transmit events generated 570 images at 2193 images/s (movie included as supplementary material).
Each frame of the RF data was beamformed using a conventional delay-sum algorithm (Fig.3(a)), producing 256 B-mode lines (i.e. 32 lines were beamformed in a parallel-receive fashion per emission) \citep{espindola2017flashfocus,shattuck1984explososcan}. The interframe displacements between the beamformed images were tracked using an adaptive algorithm validated to measure shear shock wave displacements with an accuracy of 1$ \rm \mu m$ \citep{pinton2014adaptive,Espindola2017}. The algorithm maximizes correlation values by adapting the window size based on the displacement and further, rejects poorly correlated displacement estimates using a quality-weighted median filter.

\begin{figure}[t]
\setlength{\unitlength}{\textwidth}
\begin{picture}(0.172,0.57)(0.3,0)
\put(0.4,-0.0){\includegraphics[width=0.67\textwidth,trim=0 0 0 0,clip]{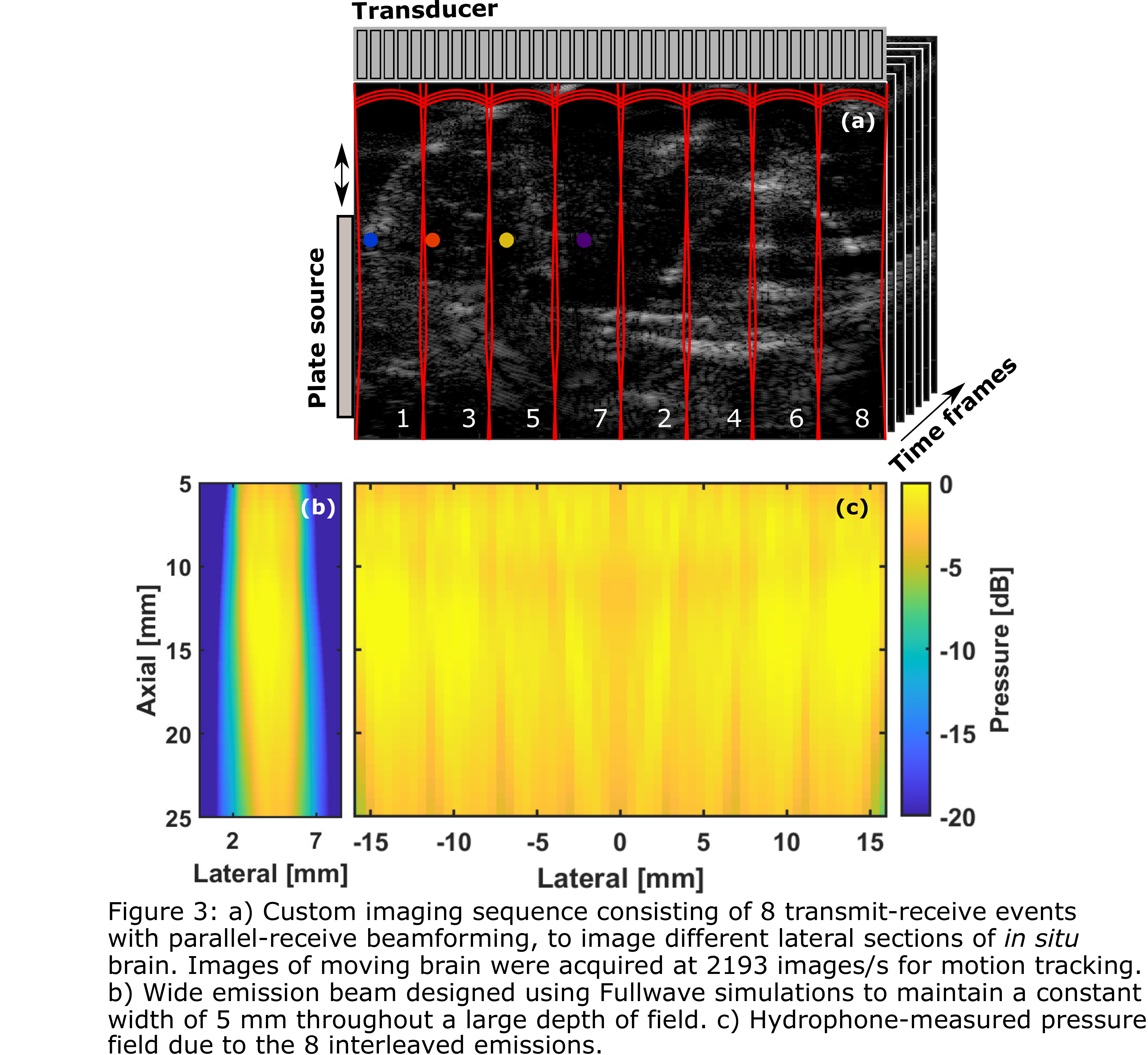}}
\put(0.36,0.48){\scriptsize \color{black}{\textcolor{white}{ \textbf{(a)}}}}
\put(0.087,0.29){\scriptsize \color{white}{\textbf{(b)}}}
\end{picture}
\end{figure}

%

As opposed to using a conventional focused emission, a custom wide-beam emission was designed with a consistent width of 5mm throughout its axial propagation (Fig.3(b)), in order to maintain a high SNR near the beam edges. The beam was designed using Fullwave ultrasound simulations by placing a 5mm wide, thin scatterer in the simulation field. By placing a receiver at the top of the acoustical simulation field, the RF data emitted from the scatterer was collected and time-reversed to produce the emission pulse used in the \textit{in situ} brain imaging sequence (Fig.3). The acoustical pressure field of the wide beam was measured in water at a transmit voltage of 4V, using a needle hydrophone with a frequency range of 1-10~MHz (HNA-0400, ONDA, Sunnyvale, California, USA), over a depth range of 5-30~mm using a sampling step-size of 500~$\rm \mu m$ laterally and axially. Fig.3(b) shows a single "pencil" beam (grid-size interpolated by a factor of 25) uniformly illuminating the area within its lateral extent, providing a long depth of field. Together, the eight beams illuminate the whole field of view (Fig.3(c)). 

\begin{figure}[htbp]
\centering
\includegraphics[width=0.69\textwidth,trim=0 0 0 0,clip]{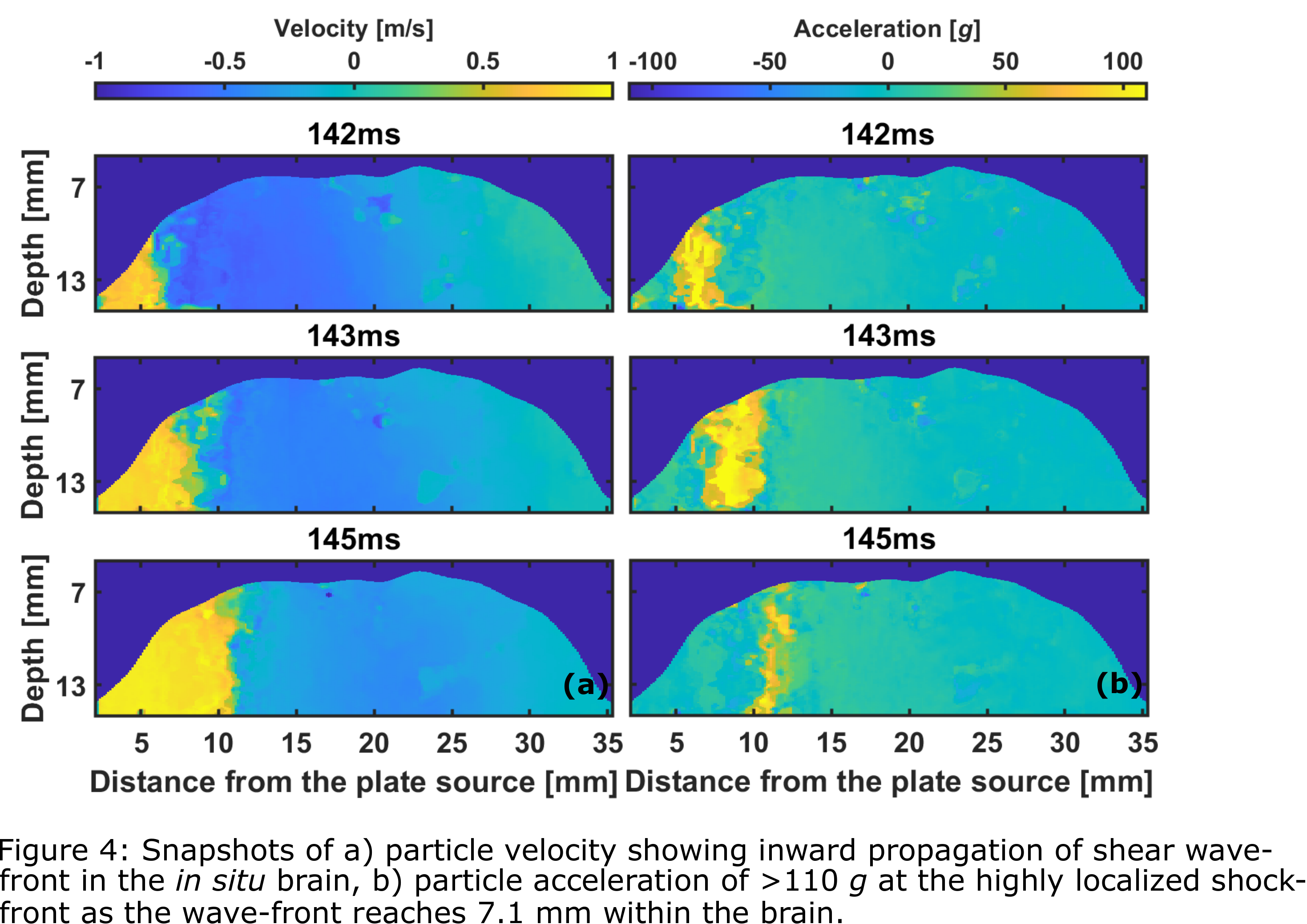}
\label{fig:snapshots}
\end{figure}

\subsection{Validation using shear shock wave simulations in brain}
Experimental results were compared to theoretical predictions, by simulating the nonlinear brain deformation due to the propagation of shear shock waves, based on the formulation and numerical methods proposed by Tripathi \textit{et al.} \citep{Tripathi2019_PPM2D}.  Briefly, a system of nine first-order balance laws was proposed to describe the propagation of linearly-polarized nonlinear shear waves in a plane orthogonal to the axis of particle displacement in a homogeneous, isotropic, relaxing soft solid. The nonlinear propagation is governed by a cubic nonlinear term resulting from a fourth-order expansion of the strain energy density function, with a nonlinear parameter $\beta = 3(\mu + A/2 +D)/2\mu$ where $A$ and $D$ are third and fourth order elastic constants, respectively \citep{Landau1986}. Since the brain is highly attenuating and dispersing, a three-body generalized Maxwell body was used to model the relaxing media i.e. the attenuation power law was $\alpha(\omega) = a\omega^b$ and the dispersion was given by the Kramers-Kronig causality conditions (more details in \citep{Tripathi2019_PPM2D}).

\begin{figure}[htbp]
\centering
\includegraphics[width=0.59\textwidth,trim=0 0 0 0,clip]{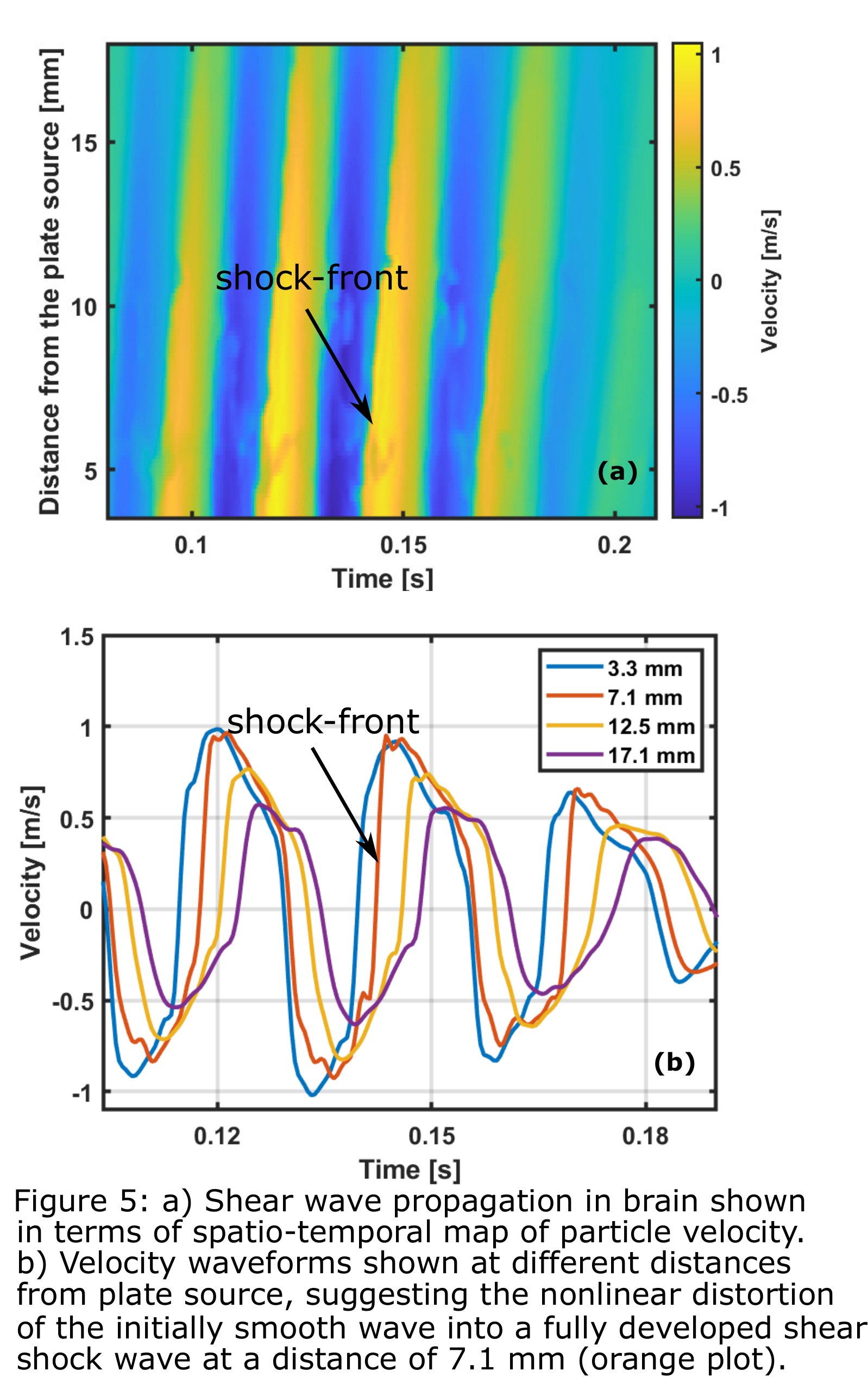}
\label{fig:velocity}
\end{figure}

The simulations were performed in a $95\times75$ mm$^2$ domain, using a circular arc (radius 26~mm, arc length 35~mm) as the source plate. The experimental data recorded near the source plate was used as the excitation pulse for the simulation, with an amplitude of 1~m/s at a frequency of 40~Hz. The spatial domain was discretized with $\Delta x = \Delta y = 130~\mu$m and $\Delta t = 23~\mu$s. The linear shear speed in the brain medium was defined as $c_T=2.10$~m/s at 75 Hz, with a density of 1000 kg/m$^3$, nonlinear parameter $\beta = 44.24$ and attenuation $0.06\omega^{1.05}$ Np/m \citep{Tripathi2020}.

\begin{figure}[t]
\centering
\includegraphics[width=0.59\textwidth,trim=0 0 0 0,clip]{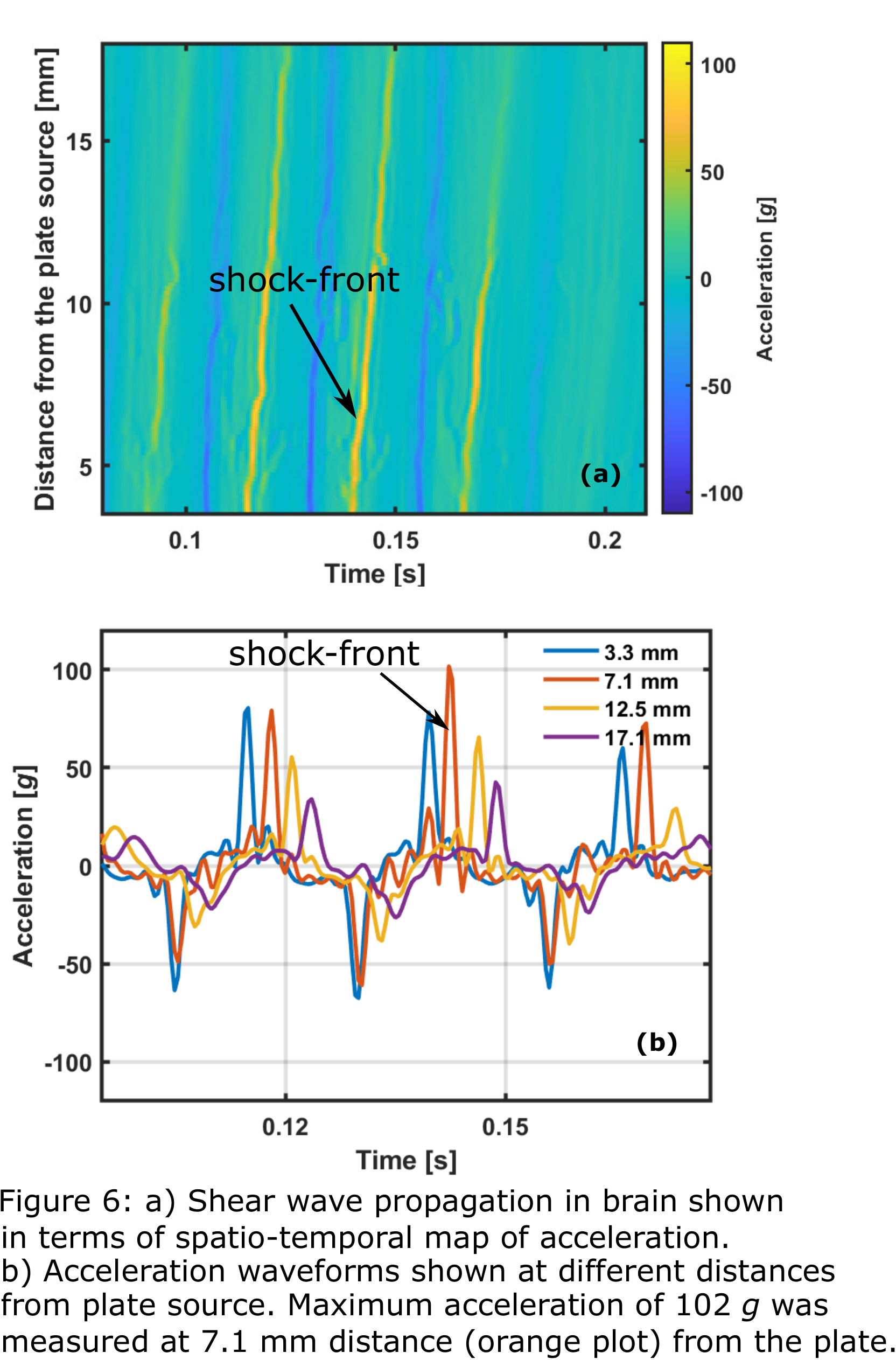}
\label{fig:accln}
\end{figure}

\section{Results}
By dividing the tracked brain displacements by the interframe sampling period the local particle velocity was calculated (Fig.4(a)). The rightward propagation of the wave-front into the brain is clearly visible in the cross-section view in Fig.4(a).

Acceleration being a more relevant metric for brain injury, was calculated using a frequency-based derivative \citep{oppenheim1997signals} as, ${a(t)=\mathcal{F}^{-1}(2{\pi}i f \mathcal{F}(v(t))})$, where $a(t)$ is the acceleration time signal, ${i=\sqrt{-1}}$, $f$ is spectral frequency and $v(t)$ is the velocity time signal. 
Numerical differentiation methods can be noise sensitive and it is critical that our acceleration estimates are accurate particularly at the shear shock-front. Noise in the experimental velocity was reduced prior to the derivative calculation using a low-pass Butterworth filter, with a cut-off frequency at the 13 \textsuperscript{th} harmonic (520~Hz for the shear wave central frequency of 40~Hz). This spectrum-based differentiation method was previously validated using a simulated known velocity signal with a similar noise level and sampling rate as experimental ultrasound-based velocity \citep{Espindola2017}. It was shown that the method preserved the shock-front acceleration peaks and underestimated the magnitude by $\sim$15\%.

\begin{figure}[htbp]
\centering
\includegraphics[width=0.89\textwidth,trim=0 0 0 0,clip]{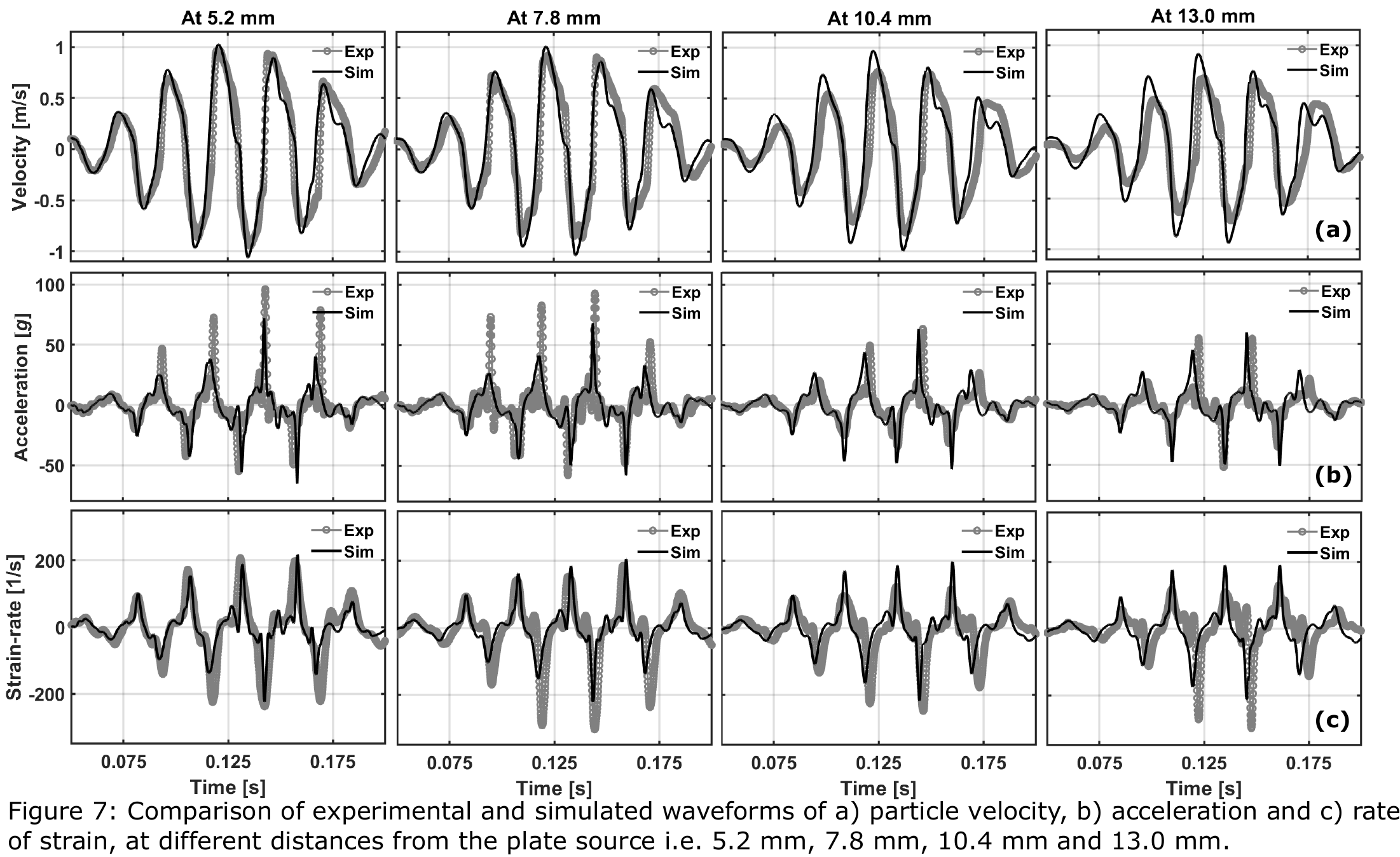}
\label{fig:sims}
\end{figure}

Snapshots of the particle acceleration movie (included as supplementary material) show that the acceleration amplitude reaches up to $>$110$g$ (Fig.4b) and that the development and decay of shock-front is a rapid event lasting $<$5ms and $<$15mm of propagation distance.

The spatio-temporal map of particle velocity is shown in Fig.5(a), by averaging the velocity estimates over a depth range of 13 to 15mm (Fig.4(a)). The shock-front is evident in Fig.5(a) as the sharp phase-change in the velocity profile at the two central positive cycles, up to $>$10mm from the plate source. From the velocity waveforms shown in Fig.5(b) at different distances from the plate source, the development of the shock-front is evident at 7.1~mm from the plate (orange plot in Fig.5(b) at the location of the orange marker in Fig.3(a)) from the sudden rise in slope at 0.11 s. In contrast, the shear wave at 3.3~mm from the source (blue plot, Fig.5(b)) has distorted much lesser from the smooth sinusoidal excitation imparted at the source. At farther distances of propagation, the velocity amplitude decreases due to the frequency-dependent attenuation within the medium. As a result, the slope at the shock-front wanes beyond 7.1 mm of propagation (yellow and purple plots, Fig.5(b)). 

The spatio-temporal map of the particle acceleration (Fig.6(a), shows the finely localized peak in-between the peak and trough of the wave, i.e. at the shock-front. The corresponding  waveforms at different distances from the source are shown in Fig.6(b). Positive acceleration at the shock-front grows and reaches a peak of 102$g$ at 7.1mm from the plate, amplified by a factor of 4.0 with respect to the surface excitation of 25$g$. This shock wave formation distance is much smaller than the shear wavelength (5.4~cm) and the shock-front is highly localized not just in time, but also spatially as seen by the peak acceleration diminishing by a factor of 1.6 between 7.1~mm (orange plot, Fig.6(b)) and 12.5~mm (yellow plot, Fig.6(b)) from the source. 

{Theoretical predictions from shear shock wave} simulations were compared to experimental results for four different distances of propagation = 5.2, 7.8, 10.4, 13.0 mm. An excellent match was seen between the experimental and simulated velocity at all the four locations (Fig.7(a)). The shark-fin shape of the cubically nonlinear waveform \citep{catheline2003observation} and the overall amplitude tracked closely. Correspondingly, the comparison between the experimental and simulated acceleration (Fig.7(b)) and strain-rate (Fig.7(c)) waveforms are shown. Both acceleration and strain-rate being rate-dependent metrics, show that the wave shape is far from monochromatic hence, demonstrating the physics of the nonlinear wave model. It is further seen that the simulations preserve the localized peaks in acceleration and strain-rate at the rise and fall of the shock-front (Fig.7(b) and (c)). The peak strain-rate at the shock-front reaches up to 250 $s^{-1}$ (Fig.7(c)). This exceeds by far previously reported brain strain-rates in concussed football players ($<100~  s^{-1}$) \citep{kleiven2007}.

%
%
%
%
%
%
%
%
%
%

%
%
%
%
%

%
%
%
%

\section{Discussion}
The shock wave formation in the brain is driven by high Mach numbers of $M=c_T/v_0=$~0.47 where, $c_T=$~2.14~m/s \citep{Espindola2017} is the shear wave speed and $v_0=$~1m/s is the amplitude particle shear velocity. The simplest nonlinear viscoelastic representation of brain motion during a linearly-polarized plane wave shear excitation can be described by a cubically nonlinear version of Burgers' equation~\citep{catheline2003observation,Espindola2017}. This type of motion has a characteristic odd harmonic signature wherein only the third, fifth, seventh, etc. multiples of the fundamental frequency are produced by the nonlinear propagation.
As the wave propagates, due to the cubic nonlinearity of the brain, the shear wave speed changes as a function of the local velocity amplitude \citep{Burgers1948}, resulting in an abrupt increase in slope at the peaks and troughs. 

The acceleration amplification of the shear wave due to nonlinearity, is almost instantaneous in the brain i.e., at 3.3 mm (blue plot), a peak acceleration of 
65$g$ was measured, while the initial condition given at the plate was just 25$g$ (Fig.6(b)). Experimentally the linear acceleration at the surface is not well captured owing to the noise in the displacement estimates at the plate boundary. However, the fully developed shock-front was clearly captured within the brain at 7.1~mm, close to the \textit{ex vivo}-measured shock formation of 6~mm, due to an 75~Hz impact. However, in the \textit{in situ} brain, a lower amplification factor of 4.0 was observed in the peak acceleration (as compared to a factor of 8.1 in the \textit{ex vivo} brain), for the same velocity amplitude of 1m/s. This disparity may be attributed to the difference in the input energy at the plate source between the two experimental scenarios. The attachment area between the plate source and the \textit{in situ} brain was $\rm \sim 6~cm^2$ in contrast to $\rm \sim 150~cm^2$ of embedded area of the plate inside the \textit{ex vivo} brain-gelatin medium. Besides larger contact area, anatomically, the spherical geometry of the skull is likely to have a strong focusing effect on the propagating shear shock waves in brain, which we are investigating using a human head gelatin phantom in a separate study \citep{tripathi2020shear}.

In Fig.6(b) it is seen that there is an asymmetry between the positive and negative cycles of the shear shock waveform. This is likely to be due to the shear wave reflections from the bone boundary underneath the brain. Furthermore, the particle motion is not strictly vertically polarized due to reflections from the boundaries of the small volume of the brain, which are unaccounted for in the shear shock wave simulations.


One of the principal guiding parameters for the design of the custom imaging sequence was measuring displacements at the maximum frequency contained in shear shock waves. The number of transmit events was chosen maintain a high frame-rate of 2193 images/s. Based on our previous experimental results in fresh porcine brain ~\citep{Espindola2017} shear wave propagation beyond frequencies of 1000 Hz are very rapidly attenuated and quickly fall below the sub-micron displacement sensitivity of ultrasound-based tracking algorithms~\citep{pinton2014adaptive}. 
For the proposed impact velocity and acceleration the chosen imaging frequency is conservatively high. Nevertheless, at larger impact magnitudes in different intended scenarios, higher sampling frequencies may be necessary to measure shock waves. Injurious impacts in the National Football League, for example, are in the 98 $\pm$ 28$g$ range with impact velocities in the 9.3 $\pm$ 1.9~m/s range \citep{pellman2003concussion}. The average Olympic boxing hit to the head generates accelerations that are 58 $\pm$ 13$g$ \citep{walilko2005biomechanics}. In vehicular impact tests, translational head accelerations of 29 to 120$g$ have been measured \citep{zhang2006brain}. The 25$g$ brain surface accelerations used here are thus comparatively low. However, even in this moderate to mild regime, the accelerations deep inside the {\it in situ} porcine brain grow rapidly. Locally, at the thin shock-front, they exceed 100$g$. The shock-front accelerations are expected to increase for the realistic, stronger impacts. This is driven by the Mach numbers which are orders of magnitude larger than in fluids or in classical solids.  This behavior was also previously observed in the \textit{ex vivo} porcine brain which demonstrates that the skull, which contains and supports the brain, does not significantly alter this type of nonlinear viscoelastodynamic behavior occurring on short and rapid spatio-temporal scales.  

\section{Conclusion}

It was shown that shear shock wave formation and propagation can be imaged \textit{in situ} in the porcine brain. Ultrasound imaging sequences were designed to reduce off-axis clutter and to record brain motion from a single impact. Time-reversed acoustical simulations using the Fullwave tool, were used to generate transmit profiles that were optimized to insonify a beamwidth of 1/8 of the imaging field of view.  Compared to previous shock wave imaging sequences, which were designed for a less challenging imaging environment, the proposed technique is better suited for \textit{in vivo} imaging because it only requires a single injurious impact, which is a more realistic model for TBI. Using quantitative motion estimation it was shown that an initially smooth monochromatic impact of 40~Hz frequency and amplitude of 25$g$ was amplified into a 102$g$ shear shock wave within 8~mm of propagation. 
In the \textit{in vivo} brain, the availability of biological response markers to axonal injury  due to the occurrence of high accelerations at the shock-front, can establish the link between brain injury and the shear shock waves. These ultrasound displacement imaging methods can also be used to measure shear wave propagation from traumatic impacts in other vulnerable organs in the body.

\end{document}